\pdfoutput=1

\documentclass[10pt,conference]{IEEEtran}

\usepackage{amsmath}
\usepackage{graphicx}
\usepackage{booktabs}
\usepackage{amssymb}
\usepackage{latexsym}	% \Diamond
\usepackage{array}		% p{}
\usepackage{multirow}
\usepackage{subfigure}
\usepackage{algorithm}
\usepackage{algpseudocode}
\usepackage{threeparttable}
\usepackage{paralist}   % compactitem
\usepackage{xspace}
\usepackage{color}
\usepackage{adjustbox}
\usepackage{balance}
\usepackage{wasysym}

\usepackage[hyphens]{url}

\newif\ifACM
\ACMfalse %IEEE

\newif\ifELS
\ELSfalse %ACM

\newif\ifPET
\PETtrue

\newif\ifCOMPAT
\COMPATfalse

\newif\ifMOST
\MOSTtrue

\newcommand{\name}{IFL\xspace}

\newcommand{\red}[1]{\textcolor[rgb]{1.00,0.00,0.00}{{#1}}\xspace}

\ifACM
\newcommand{\myfig}{Figure\xspace}
\else
\newcommand{\myfig}{Fig.\xspace}
\fi

\newcommand\y{$\checkmark$\xspace}
\newcommand\x{\textcolor[rgb]{1.00,0.00,0.00}{$\times$}\xspace}

\newcommand{\file}{\texttt{file://}\xspace}
\newcommand{\content}{\texttt{content://}\xspace}
\newcommand{\intent}{\texttt{intent://}\xspace}

\newcommand{\Sfa}{\texttt{SOPf1}\xspace}
\newcommand{\Sfb}{\texttt{SOPf2}\xspace}

\newcommand{\sopPFL}{\texttt{sopIFL}\xspace}
\newcommand{\aimPFL}{\texttt{aimIFL}\xspace}
\newcommand{\cmdPFL}{\texttt{cmdIFL}\xspace}
\newcommand{\serPFL}{\texttt{serverIFL}\xspace}

\newcommand{\aimPFLa}{\texttt{aimIFL-1}\xspace}
\newcommand{\aimPFLb}{\texttt{aimIFL-2}\xspace}

\begin{document}

\title{Cross-Platform Analysis of Indirect File Leaks in Mobile Applications on Android and iOS}
\title{Cross-Platform Analysis of Indirect File Leaks\\in Android and iOS Applications}

%\author{
%\IEEEauthorblockN{Anonymous Submission}
%}
\author{
\IEEEauthorblockN{Daoyuan Wu and Rocky K. C. Chang}
\IEEEauthorblockA{Department of Computing, The Hong Kong Polytechnic University \\
\{csdwu, csrchang\}@comp.polyu.edu.hk}
\red{\small This paper was published in IEEE Mobile Security Technologies 2015~\cite{MoST15} with the original title of ``Indirect File Leaks in Mobile Applications''.}
}

\maketitle

\begin{abstract}

Today, much of our sensitive information is stored inside mobile applications (apps), such as the browsing histories and chatting logs. To safeguard these privacy files, modern mobile systems, notably Android and iOS, use sandboxes to isolate apps' file zones from one another. However, we show in this paper that these private files can still be leaked by \textit{indirectly} exploiting components that are trusted by the victim apps. In particular, we devise new \textit{indirect file leak} (\name) attacks that exploit browser interfaces, command interpreters, and embedded app servers to leak data from very popular apps, such as Evernote and QQ. Unlike the previous attacks, we demonstrate that these IFLs can affect both Android and iOS. Moreover, our IFL methods allow an adversary to launch the attacks remotely, without implanting malicious apps in victim's smartphones. We finally compare the impacts of four different types of IFL attacks on Android and iOS, and propose several mitigation methods.

\end{abstract}

\section{Introduction}
\label{sec:intro}

Mobile applications (apps) are gaining significant popularity in today's mobile cloud computing era \cite{AppDaily12, AppMore14}.
%Popular web services are also experiencing huge traffic volumes from their mobile apps \cite{MobileTraffic, FBYoutube}.
Much sensitive user information is now stored inside mobile apps (on mobile devices), such as Facebook authentication tokens, Chrome browsing histories, and Whatsapp chatting logs.
To safeguard these privacy files, modern mobile systems, notably Android and iOS, use sandboxes to isolate apps' file zones from one another.
%(see Section \ref{sec:backg}).

\begin{figure}[t!]
\begin{center}
\includegraphics[width=0.36\textwidth]{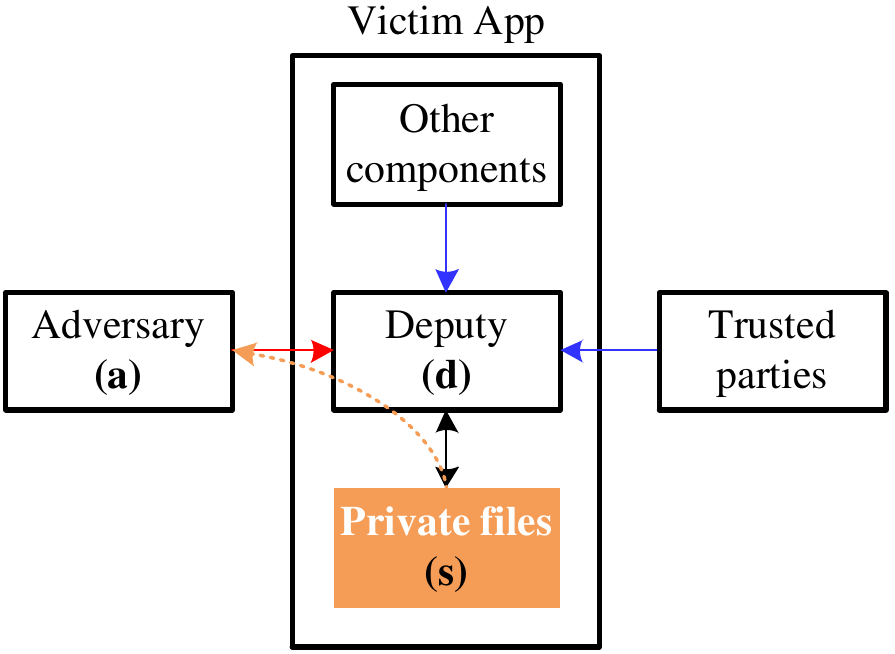}
\end{center}
\ifMOST
\vspace{-2ex}
\else
\vspace{-4ex}
\fi
\caption{A high-level \name model.}
\ifMOST
\vspace{-4ex}
\else
\vspace{-2ex}
\fi
\label{fig:overview}
\end{figure}

However, it is still possible for an adversary to steal private app files in an \textit{indirect} manner by exploiting components that are trusted by the victim apps.
We refer to this class of attacks as \textit{indirect file leaks} (IFLs).
\myfig \ref{fig:overview} illustrates a high-level \name model.
Initially, an adversary cannot directly access a private file, formulated as $a \nLeftarrow s$.
If the adversary can send crafted inputs to a deputy\footnote{We borrow this term from the classic confused deputy problem \cite{Confused88} to represent a trusted component in victim apps.} inside the victim app ($a \rightarrow d$) and these inputs can trigger the deputy to send the private file to the adversary ($d \rightarrow s \Rightarrow a$), then the adversary can indirectly steal the private file.
The whole process, $a \rightarrow d \rightarrow s \Rightarrow a$, achieves the goal of $s \Rightarrow a$, causing an \name.

In this paper, we devise new \name attacks that exploit browser interfaces, command interpreters, and embedded app servers to leak files from very popular apps, such as Evernote and Tencent QQ.
Unlike prior works \cite{ContentScope13, FileCross14} that only show local \name attacks on Android, we demonstrate that three out of our four IFL attacks affect both Android and iOS.
%, because these deputies are available on both platforms.
We summarize these attacks below.
\begin{itemize}

  \item \textit{\sopPFL attacks} bypass the same-origin policy (SOP), which is enforced to protect resources originating from different \textit{origins} (i.e., the combination of scheme, domain, and port), to steal private files via browsing interface deputies.
    Although our prior work \cite{FileCross14} has demonstrated such attacks on Android by exploiting numerous Android browsers' SOP flaws on the \file scheme, we are extending it in this paper to a number of vulnerable iOS apps, such as the very popular Evernote, Tencent QQ, and Mail.Ru.
    We also confirm that the latest iOS~8 fails to enforce the appropriate SOP on \file.
    In our analysis, the root case of this attack is that the legacy web SOP is found to be inadequate for the local schemes, such as \file. Eradicating the problem may call for an enhanced SOP.

  \item \textit{\aimPFL attacks} also leverage browsing interfaces as deputies, but they do not need to violate SOP.
    %This makes \aimPFL more powerful than \sopPFL attacks.
    It can do so by injecting and executing unauthorized JavaScripts \textit{directly} on target files, instead of requiring a malicious file to bypass SOP to access target files as in the \sopPFL attacks.
    Popular Android browsers, such as Baidu, Yandex, and 360 Browser, can be easily compromised in this way, allowing their private files (e.g., cookie and browsing history) to be stolen.
    %Besides 360 Mobile Safe and Baidu Search, we also uncover a vulnerable iOS app, myVault.
    The high-profile 360 Mobile Safe and Baidu Search are also exploitable.
    Besides these Android apps, we further uncover a vulnerable iOS app, myVault.

  \item \textit{\cmdPFL attacks} exploit command interpreters as deputies inside victim apps to execute unauthorized commands for file leaks.
    We demonstrate that the top command apps on Android, Terminal Emulator and SSHDroid, can be stealthily exploited to execute arbitrary commands, possibly with the root privilege.
    This will jeopardize their own files (e.g., command histories and private configuration files), sensitive user photos stored on SD cards, and even other apps' private files.

  \item \textit{\serPFL attacks} send unauthorized file extraction requests to embedded app server deputies inside victim apps to obtain private files.
    All of the tested popular server apps, such as WiFi File Transfer (Android) and Simple Transfer (iOS), are vulnerable to these attacks. %in an Intranet
    It is worth noting that both the \cmdPFL and \serPFL attacks use previously unexplored deputies---command interpreters and embedded app servers---to launch \name attacks for the first time in mobile platforms.

\end{itemize}

Besides the cross-platform vulnerability, our \name methods also allow an adversary to launch the attacks remotely, without implanting malicious apps in victim's phones as in prior works \cite{ContentScope13, FileCross14}.
These \name attacks can be launched both locally (in the same phone) and remotely (in an Intranet/Internet).
Table \ref{tab:main} highlights the identified \name vulnerabilities and their major attack channels.
Besides local \name attacks, we show that browsers, such as Baidu and Yandex Browser, can be remotely exploited by enticing the victim to access a web page.
Email apps (e.g., Mail.Ru) and social apps (e.g., QQ) can be similarly compromised if the victim opens a malicious attachment or file transmitted by a remote adversary. In other remote attacks, the adversary can scan the whole Intranet, locate open ports, and exploit vulnerable server apps installed on the victim phone.
%We will show how weak the authentication mechanisms are used by these apps;

Furthermore, we analyze the differences between Android and iOS in terms of the impact of the \name attacks.
These differences are caused by different system architectures and app design practices between Android and iOS.
Our analysis shows that a common iOS app practice could lead to more powerful and pervasive \sopPFL attacks on iOS than Android. On the other hand, three iOS system characteristics help lessen the impacts on iOS for the other three \name attacks. These findings can help developers and OS providers build more secure apps and mobile systems.

\textbf{Ethical considerations.}
All of our vulnerability testing is conducted using our own mobile devices and test accounts.
The tests \textit{never} affect the data security of real-world users.
As the \name attacks in mobile apps are client-side vulnerabilities, they \textit{cannot} affect the server-side integrity.

\textbf{Real-world impacts.}
We have reported most of the identified vulnerabilities to their vendors in a responsible way and are in the process of reporting the remaining vulnerabilities.
All of the contacted vendors have acknowledged our reports. For example, Evernote has listed us in its security hall of fame.
Baidu has ranked one of our reports as the most valuable vulnerability report of the second quarter of 2014,
and Qihoo 360 has issued us the highest award in its mobile bug bounty program history.
We have also offered our suggestions to the vendors to fix the identified vulnerabilities.
%For example, two vulnerable Mail.Ru apps have been patched.

\begin{table}[t!]
\centering
\vspace{-2ex}
\caption{Four \name vulnerabilities and their attack channels.}
\vspace{-2ex}
\scalebox{
\ifACM
\ifELS
0.76
\else
0.9
\fi
\else
1
\fi
}{
\begin{adjustbox}{center}
\begin{tabular}{ c | c | c| }

\cline{2-3}
  & \multicolumn{2}{c|}{Attack Channels} \tabularnewline
\cline{2-3}
  & Remote & Local \tabularnewline
\hline

\multicolumn{1}{|c|}{\multirow{2}{*}{\sopPFL}}  & Evernote, Tencent QQ  & UC \& QQ browsers \tabularnewline
\multicolumn{1}{|c|}{}                          & Mail.Ru               & 360 \& Mail.Ru Cloud \tabularnewline
\hline

\multicolumn{1}{|c|}{\multirow{2}{*}{\aimPFL}}  & \multicolumn{2}{c|}{Baidu, Yandex, and 360 browsers} \tabularnewline
\multicolumn{1}{|c|}{}                          & \multicolumn{2}{c|}{360 Mobile Safe, Baidu Search, myVault} \tabularnewline
\hline

\multicolumn{1}{|c|}{\cmdPFL}                   & SSHDroid              & Terminal Emulator \tabularnewline
\hline

\multicolumn{1}{|c|}{\serPFL} & \multicolumn{2}{c|}{WiFi File Transfer, Simple Transfer} \tabularnewline
\hline
\end{tabular}
\end{adjustbox}
}
\ifMOST
\vspace{-2ex}
\fi
\label{tab:main}
\end{table}

\textbf{Contributions.}
To summarize, we make the following three contributions:
\begin{compactitem}

\item We devise four new \name attacks that, for the first time, can affect both Android and iOS and are exploitable not only locally but also remotely. (Section \ref{sec:problem} \& \ref{sec:evaluate})

\item We identify a number of zero-day \name vulnerabilities in popular Android and iOS apps and uncover a serious SOP issue in the latest iOS 8 system. (Section \ref{sec:evaluate})

\item We analyze the differences between Android and iOS in terms of the \name attacks' impacts and propose several methods to mitigate the attacks. (Section \ref{sec:implication} \& \ref{sec:mitigate})

\end{compactitem}

\section{Background}
\label{sec:backg}

\subsection{Sandbox-based App Isolation}

Both Android and iOS use sandbox-based app isolation to build a trustworthy mobile environment.
Each app resides in its own sandbox, with its code and data isolated from other apps.
This isolation is usually enforced at the kernel level.
For example, Android uses UNIX UID-based protection to isolate each app, in which each app is treated as an independent user and runs in a separate process with a unique \texttt{uid}.

Each app's sensitive files are stored in their own system-provided isolated (or private) file zone.
Unless an app actively leaks a file (i.e., a direct file leak), other apps have no access to the protected files.
The widely deployed SEAndroid MAC system \cite{SEAndroid13} further thwarts the risks incurred by direct file leaks.
However, IFLs can still occur in the presence of both sandbox-based isolation and MAC. Although the actual executer of the file access is the legal victim app, the file request is actually initiated and crafted by an adversary.
Encryption-based defenses, such as encrypting all private app files, face similar limitations as the MAC. To sum up,
\name remains a serious, and yet unsolved, threat, which motivates our study.

\subsection{Terminology}
In Table \ref{tab:term}, we summarize the terms used throughout this paper.

\ifMOST
\begin{table}[ht!]
\else
\begin{table}[t!]
\fi
\centering
\vspace{-4ex}
\caption{Terms and their descriptions.}
\scalebox{
\ifACM
\ifELS
0.76
\else %PET
0.82
\fi
\else %IEEE
0.94
\fi
}{
\begin{adjustbox}{center}
\begin{tabular}{ |c | l| }

\hline
Term  &  Description \tabularnewline
\hline
\hline

\multirow{2}{*}{Private files}  &  The files stored in apps' isolated file zones. In nonlocal \name  \tabularnewline
                &  attacks, they also include files on a SD card, e.g., user photos.  \tabularnewline
\hline
Target file   &  A private file the adversary wants to steal in an \name attack.  \tabularnewline
\hline
\multirow{2}{*}{Permission}     &  A form of privilege representation. For example, the \texttt{INTERNET}  \tabularnewline
                &  permission on Android and the \texttt{Contact} permission on iOS.  \tabularnewline
\hline
\multirow{2}{*}{Root}           &  A superuser privilege. For example, a \textit{rooted phone} is a phone \tabularnewline
                &  that enables superuser privilege for apps.  \tabularnewline
\hline
Browsing        &  Or browsing component. A component with browsing capabili-  \tabularnewline
interface       &  ty, usually built with Android's WebView or iOS's UIWebView.  \tabularnewline
\hline
Command         &  \multirow{2}{*}{An app component that can interpret and execute commands.} \tabularnewline
interpreter     &  \tabularnewline
\hline
App server      &  A server component embedded in an app.  \tabularnewline

\hline
\end{tabular}
\end{adjustbox}
}
%\vspace{-4ex}
\label{tab:term}
\end{table}

\section{The \name Attacks}
\label{sec:problem}

In this section, we first describe the adversary model and then detail the four types of \name attacks introduced in Section~\ref{sec:intro}.
%In this section, we first generalize Indirect File Leak (\name) attacks, and then propose three new \name attacks.
%We also give a deeper understanding of a known yet important \name attack in Section \ref{sec:sopAtk}.

\subsection{Adversary Model}
\label{sec:threatmodel}

We consider the following three types of adversaries in our \name attacks. A local adversary can launch only local attacks, whereas the Intranet and Internet adversaries \textit{remote} \name attacks.
\begin{itemize}

  \item A \textit{local} adversary is an attack app installed on the same smartphone as the victim app.
    It requires few or no permissions and does not exhibit any typical malicious app behavior \cite{AndroidMalware12}.
    The root privilege is \textit{never} used by this attack app.
    %So users cannot differentiate them with other apps.
    We also do not consider screenshot attacks~\cite{Screenmilker14} that require strong assumptions.
    %\TODO local \name attacks

  \item An \textit{Intranet} adversary resides in the same Intranet as the victim's mobile device.
    It can send network requests to any other node within the Intranet.
    It can sniff the nearby wireless traffic and retrieve unencrypted content.
    We do not assume that it can launch effective ARP spoofing attacks, as network administrators can detect such anomalous events.

  \item An \textit{Internet} adversary can be located in any host in the Internet.
    It remotely compromises a victim by (i) enticing a victim to browse a web page under the adversary's control, and/or (ii) sending the victim a malicious file via email, chatting app, social network, and other means.

\end{itemize}

\subsection{Bypassing SOP on Browsing Interfaces}
\label{sec:sopAtk}

%We review and complement a previously reported \name attack, which we call \sopPFL, to obtain a deeper understanding based on the prior work in~\cite{FileCross14}.

The \sopPFL attacks bypass SOP to steal private files via browsing interfaces.
The deputy in this attack is the browsing interface or rendering engine, whose SOP enforcement is flawed and cannot prevent malicious JavaScript codes from accessing a private file.
All apps that contain browsing components are potential victims.

The \file scheme is an ideal medium to launch the \sopPFL attacks.
Two parts of SOP enforcement on \file can be exploited to steal local private files.
The adversary can cross the origin from a web domain to access local file content, if the cross-scheme SOP enforcement for \texttt{http(s)://} to \file (labelled as \Sfa) is broken.
Alternatively, the adversary can leverage a local malicious HTML file in one path to steal a target file in another path.
In this case, the \file SOP enforcement between the two file origins (labelled as \Sfb) must be bypassed.
Since failure of enforcing \Sfa is rare in modern rendering engines, we focus on the \sopPFL attacks that will bypass the \Sfb.

\ifMOST
\else
Our recent study~\cite{FileCross14} shows that Android does not effectively enforce \Sfb.
Android prior to 4.1 does not enforce this policy at all.
Although the succeeding Android versions fix this logical flaw at the engine level, a new application program interface (API)~\cite{setAllowFileAccessFromFileURLs} is available for developers to loosen the corresponding SOP restriction.
% \TODO API prevalence \TODO
An app compiled with a vulnerable software development kit (i.e., before 4.1) is still exploitable on more recent Android platforms, including Android 4.4, which uses the Chrome Blink engine as its default engine.
% \TODO Even worse, an implementation flaw using
\fi

\ifMOST
Our recent study~\cite{FileCross14} shows that Android does not effectively enforce \Sfb.
\fi
However, little is known about iOS.
This is where our contribution for the \sopPFL attacks lies.
Contrary to our expectation, these attacks can have higher impact on the iOS ecosystem than the Android's.
Our testing using iPhone 6 reveals that even the latest iOS 8 does not properly enforce \Sfb.
Indeed, iOS \textit{never} guarantees this policy (Section \ref{sec:sopPFL}).
In Section \ref{sec:implication}, we identify a common practice among the iOS apps that could lead to more pervasive and powerful \sopPFL attacks on iOS than Android.

We believe that the root problem is that the legacy SOP cannot adequately cover the local schemes, such as \file.
The typical web SOP principle (i.e., the legacy SOP) allows file A (at \path{file:///dir1/a.html}) to access file B (at \path{file:///dir2/b.txt}) because the two origins share the same scheme, domain (i.e., 127.0.0.1 or localhost), and port.
In practice, this legal behavior fails to meet the security requirements for \file, especially in the mobile environment.
Therefore, an enhanced SOP for local schemes, such as adding the ``path'' element to the current three-element SOP tuples, is needed for eradicating this vulnerability.
%This recommendation is necessary and urgent, according to our vulnerability testing results presented in Section \ref{sec:sopPFL}.
We reported our iOS findings to Apple on 19 January 2015 and suggested them to use an enhanced \file SOP at the system or engine level.
%They are currently handling it.

%symbolic link attack
%desktop browser file SOP

%\TODO Other local schemes include ...

%\TODO Accessing private file zones is the root source for many attacks.

%=============
%  aim PFL
%=============
\subsection{Unauthorized JavaScript Execution on Target Files}
\label{sec:aimAtk}

The \aimPFL attacks could be regarded as an advanced variant of the \sopPFL attacks.
Both attacks use the browsing interface as the deputy, but the \aimPFL attack does not violate SOP.
It can do so by injecting and executing unauthorized JavaScripts \textit{directly} on target files, instead of requiring a malicious file to bypass SOP to access target files as in the \sopPFL attacks.

The \aimPFL attacks usually consist of two steps.
Unauthorized JavaScript codes are first \textit{injected} into the target file.
This can be achieved in different ways, depending on the type of the target files.
For example, if the target is a cookie file, the JavaScript codes can be injected via a website's cookie field.
The target files are then \textit{loaded} and rendered.
The previously injected JavaScript is executed to steal the current file content via an HTML document object model variable, such as \texttt{document.body.innerHTML}.
As JavaScript only accesses the current document, this attack does not violate SOP, thus setting this attack apart from the \sopPFL attack.

It is worth noting that the two steps can also be performed simultaneously without the user's knowledge.
A victim user's private files will be stolen when he browses a web page under the attacker's control.
In this paper, we focus on designing remote \aimPFL attacks, although they can also be conducted locally.

We identify two types of remote \aimPFL attacks illustrated in \myfig \ref{fig:aimPFL}.
In the first type, a web page tries to load a target file through local schemes like \file and \texttt{content://}.
The file can be loaded by a file link or an HTML \texttt{iframe}.
The link-based loading requires an extra user clicking, whereas the iframe-based loading is automatic and does not require any user action.
%The web page normally cannot access the cookie file content, because the correct SOP restricts the cross from \texttt{http://} to \file and \texttt{content://}.
Before loading the target cookie file, the web page injects malicious JavaScript codes (e.g., \path{<script>alert(document.body.innerHTML)</script>}) via the web cookie field.
Once the cookie file is successfully loaded, the JavaScript inside it can steal the cookie content.
As will be shown in Section \ref{sec:aimPFL}, the popular 360 Mobile Safe, Baidu, and Yandex browsers are all vulnerable to this type of \aimPFL attack.

\begin{figure}[t!]
  \centering
  \vspace{-2ex}
\begin{adjustbox}{center}
  \subfigure[\small Attacks that actively load a target file.] {
	\label{fig:aimPFL1}
    \ifMOST
    \includegraphics[height = 42ex]{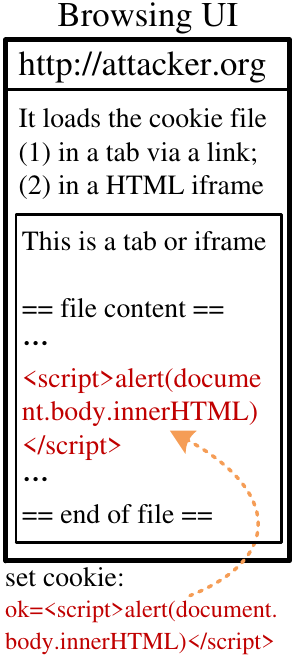}
    \else
    \includegraphics[height = 50ex]{aimPFL1}
    \fi
  }
  \ifMOST
  \hspace{3ex}
  \else
  \hspace{6ex}
  \fi
  \subfigure[\small Attacks that exploit victim apps' file loading features.] {
	\label{fig:aimPFL2}
    \ifMOST
    \includegraphics[height = 42ex]{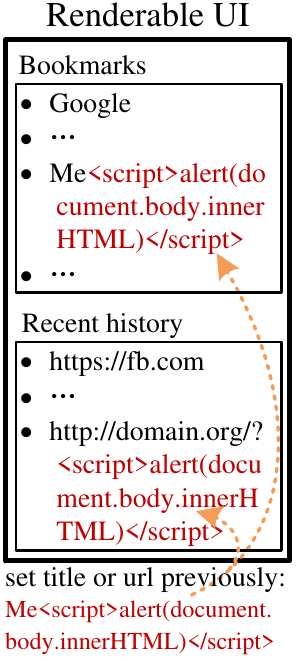}
    \else
    \includegraphics[height = 50ex]{aimPFL2}
    \fi
  }
\end{adjustbox}
\vspace{-2ex}
\caption{Remote \aimPFL attacks.}
\vspace{-4ex}
\label{fig:aimPFL}
\end{figure}

The second type of the \aimPFL attacks does not actively load and render target files.
Instead, it exploits victim apps' ability to load the content of target files into a renderable user interface (UI), such as those containing WebView widgets.
For example, some browser apps load browsing histories from the history file into a renderable UI.
Using this app-loading feature, an adversary simply injects the JavaScript into the target file.
For example, as illustrated in \myfig \ref{fig:aimPFL2}, the adversary injects an unauthorized JavaScript into the history file through the title or URL field of a web page.
When the user switches to the renderable UI, a new history log containing the malicious title or URL is displayed.
The embedded JavaScript is then executed to steal the history file.
These passively loaded files are usually rendered in the local domain (e.g., under \file).
The adversary can also combine an \aimPFL attack with a \sopPFL attack to steal other private files that are not loaded by the victim apps.
We will detail the affected apps in Section \ref{sec:aimPFL}.

%=============
%  cmd PFL
%=============
\subsection{Unauthorized Command Execution on Command Interpreters}
\label{sec:cmdAtk}

The \cmdPFL attacks exploit command interpreters (as deputies) inside victim apps to execute file operation related commands for \name attacks.
We consider explicitly embedded command interpreters, such as those in command terminal apps.
In other words, (remote) code execution vulnerabilities contained in host apps are out of the scope.
%We call the first type of \cmdPFL attacks \ecmdPFL and the second \icmdPFL, to simplify the following description.

An app is vulnerable to a \cmdPFL attack only when it can be injected with unauthorized commands.
To leak private files, the injected commands can
(i) set a world-readable file permission by invoking the \texttt{chmod} command,
(ii) export a file to a public SD card via the \texttt{cp} command,
or (iii) send a file to a remote server through commands like \texttt{scp}.
%(4) even more stealthily, opening and displaying a file to launch screenshot attacks~\cite{Screenmilker14},
%and etc.
If the victim app has root permissions, all of these commands can be used to steal private files in other, possibly more sensitive, apps.
% \TODO (3) is client, while the fourth app is the server

%An app is vulnerable to a \cmdPFL attack only when it can be injected with unauthorized commands.
%This is straightforward for \icmdPFL vulnerabilities, because remote code execution (RCE) flaws inside them allow an adversary to issue commands directly.
%Therefore, identifying an \icmdPFL vulnerability simply requires determining whether an app has RCE flaws.
%RCE flaws can be caused by API design errors (e.g., the Android addJavascriptInterface issue \cite{addJavascriptInterfaceSaga}) or classic memory corruption vulnerabilities (e.g., an integer overflow was used to exploit Chrome for Android in Mobile Pwn2Own \cite{pwn2own13}).
%Detecting general RCE flaws is beyond the scope of this paper, but we uncover several addJavascriptInterface issues in popular apps (Section \ref{sec:cmdPFL}).

To discover and exploit a \cmdPFL vulnerability, an adversary needs to identify channels that can be used to confuse command interpreters to accept unauthorized commands.
The cross-app component communication channel \cite{ISC10_Privilege, ComDroid11} on Android can be used to launch local \cmdPFL attacks, if the command interpreter components are exposed.
\ifMOST
An adversary can also exploit the URL scheme \cite{CrossOrigin13} to achieve the same, if not better, attack impacts.
\else
Likewise, the URL scheme \cite{CrossOrigin13, IntentScheme14, AutomateiOS, iPhoneURL} can achieve the same, if not better, attack effects.
\fi
Moreover, general remote \cmdPFL attacks are also possible, if the command interpreter accepts remote command requests. In addition, victim apps' configurations stored in public storage can be changed to indirectly inject commands. Accessibility services can also be misused to mimic user commands \cite{A11y14}.

%\TODO one problem: the private files in these apps are not very sensitive. But the photos are in the public SD card.

A \cmdPFL attack could also be launched through a GUI to attack file manager apps.
An adversary can force these apps to perform unauthorized UI-based file operations to leak file contents.
However, this is not a real threat, because such file operations have to be conducted by victim users and thus are easily noticed.
A smarter adversary can wait for users to open a sensitive file and launch screenshot attacks \cite{Screenmilker14} to sniff the content. 
We exclude this scenario also from our treat model, because this requires the adversary to closely monitor the victim's activity.

%=============
%  server PFL
%=============
\subsection{Unauthorized File Extraction via Embedded App Servers}
\label{sec:serAtk}

The confused deputies here are the app servers embedded in victim apps.
An adversary sends unauthorized file extraction requests to exploit vulnerable embedded app servers for obtaining private files.
The affected app servers are mainly file servers (e.g., those that support \texttt{http://} and \texttt{ftp://} requests) that provide users with file transmission service between phones and desktops.
The command servers mentioned in the last section are another type of candidate servers.
%Traditional web servers can also be embedded in apps, allowing \serPFL attacks based on path traversal \cite{JavaVulnerability05}.
Some apps that support multi-function servers are also affected, such as the very popular AirDroid app, which has over 10 million installs.

\ifMOST
A \serPFL attack can be conducted in three ways.
\else
A \serPFL attack can be conducted in three ways, as shown in \myfig \ref{fig:serPFL}.
\fi
First, a local attack can be launched from another app installed on the same smartphone as the victim app.
It scans the local hosts' ports and sends packets to the port listened to by the victim app.
Second, adversarial nodes (e.g., phones or laptops) in the same Intranet can attack the victim app in another device.
Within the same Intranet, it is also easy for an adversary to identify a port opened by the victim app.
Third, we show that remote \serPFL attacks are also possible.
Attack vectors are delivered through the victim's desktop browsers when they browse a malicious web page.

\ifMOST
\else
\begin{figure}[t!]
\begin{center}
\includegraphics[width=0.4\textwidth]{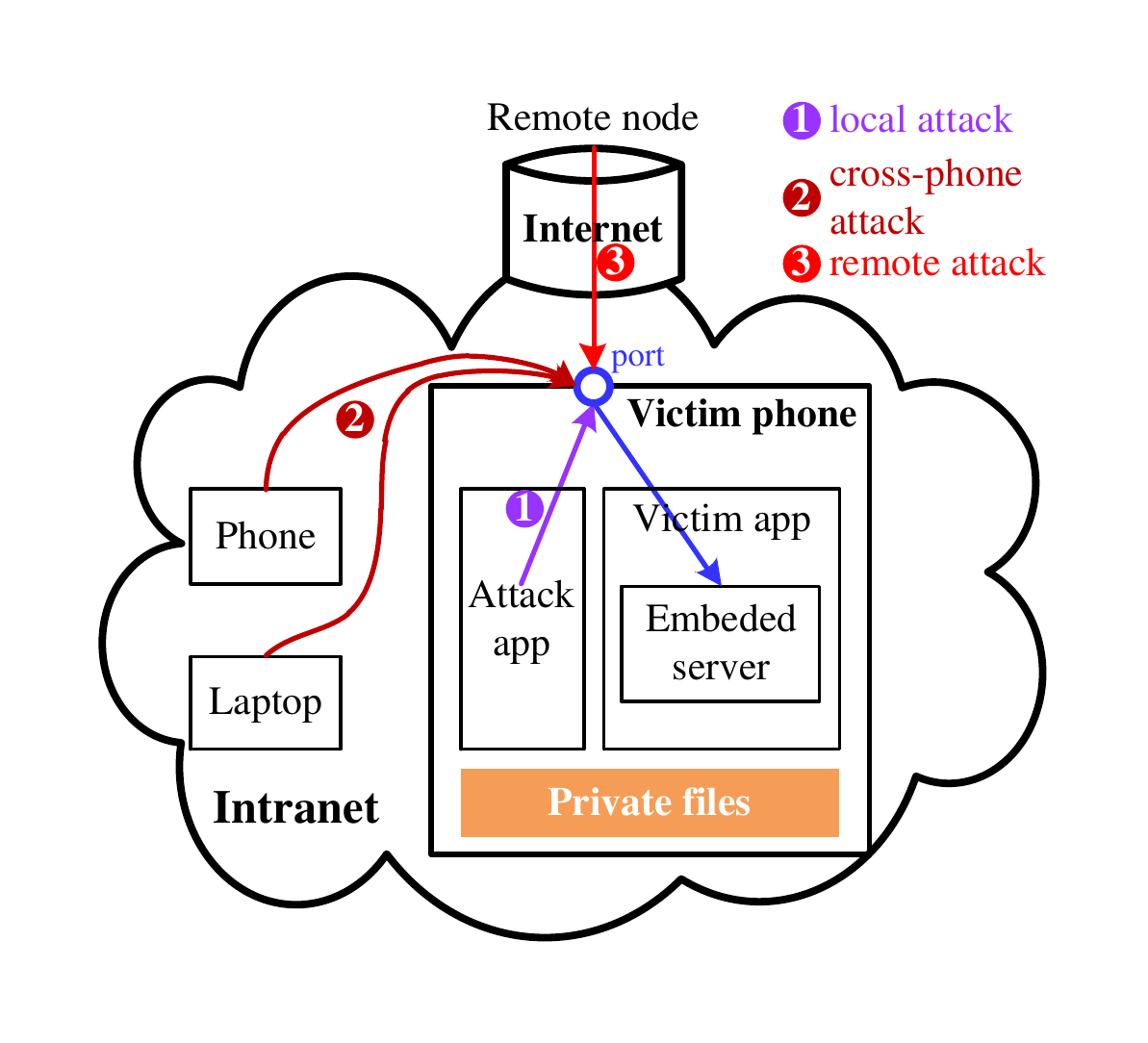}
\end{center}
\vspace{-4ex}
\caption{Three kinds of \serPFL attacks.}
\vspace{-3ex}
\label{fig:serPFL}
\end{figure}
\fi

A successful \serPFL attack may need to bypass authentication set up to protect the victim app, such as an authentication code (e.g., user password) or a confirmation action (e.g., clicking a confirmation button).
Our evaluation in Section \ref{sec:serPFL}, however, shows that the current authentication in server apps is nearly broken.
Some apps use no or weak authentication (e.g., using only four random numbers).
Almost all channels for transmitting authentication codes and subsequent session tokens are unencrypted.
An Intranet adversary can easily sniff these secrets to bypass the authentication step.
%Finally, \ldots \TODO

%=============
%  Summary
%=============
%\subsection{TODO}

%display the file, so that it can be.
%root file manager (the with GUI version of commander)

%\TODO assess each attack's severity. Or put this in the next section

\section{Uncovering \name Vulnerabilities in Android and iOS Apps}
\label{sec:evaluate}

\ifPET
\else
\input{method}
\fi

%=============
%  sop PFL
%=============
\subsection{SopIFL Vulnerabilities}
\label{sec:sopPFL}

\ifMOST
\else
%This section describes identified \sopPFL vulnerabilities to support the new understanding in Section \ref{sec:sopAtk}.

We extend the prior work \cite{FileCross14} by demonstrating that the \sopPFL vulnerabilities also exist in non-browser Android apps.
Specifically, we discover two vulnerable Android apps, 360 Cloud and On The Road (a Chinese travel app).
Theses apps are not browsers, but they contain exposed browsing interfaces that can be exploited by local \sopPFL attacks.
% \TODO demonstrate the boundary of browsers is not clear in the mobile platform
\fi

\ifMOST
We uncover a number of vulnerable iOS apps summarized in Table \ref{tab:iOSresults}.
\else
Our main result is uncovering a number of vulnerable iOS apps.
\fi
We evaluate them using an iPhone 6 (with the latest iOS 8) and an iPad 3 (with iOS~7).
Our evaluation shows that both iOS 7 and 8---which are used in around 95 percent of all iOS devices~\cite{iOSmarket}---do not enforce \Sfb at the engine level. Since all iOS apps can use only the default web engine, the current solution for mitigating the \sopPFL vulnerabilities can only be done on the application level.

\begin{table}[t!]
\centering
\vspace{-3ex}
\caption{\small iOS apps vulnerable to the \sopPFL attacks.}
\ifMOST
\vspace{-2ex}
\fi
\scalebox{
\ifACM
\ifELS
0.9
\else
0.94
\fi
\else
1
\fi
}{
\begin{adjustbox}{center}
\begin{tabular}{ |c | c | c| }
\hline
Category & Vulnerable Apps & Attack Channel \tabularnewline
\hline
\hline
\multirow{2}{*}{Browser} & UC, Mercury      & \multirow{2}{*}{Local} \tabularnewline
 	                     & Baidu, Sogou, QQ browsers &  \tabularnewline
\hline
\multirow{2}{*}{Cloud Drive} & Mail.Ru Cloud      & \multirow{2}{*}{Local \& Web} \tabularnewline
 	                         & Baidu Cloud, 360 Cloud &  \tabularnewline
\hline
Note/Read & Evernote, QQ Reader & Local \& Web \tabularnewline
\hline
Email & Mail.Ru & Remote \tabularnewline
\hline
Social & Tencent QQ & Remote \tabularnewline
\hline
Utility & Foxit Reader, OliveOffice & Local \tabularnewline
\hline
\end{tabular}
\end{adjustbox}
}
\ifMOST
\vspace{-4ex}
\fi
\label{tab:iOSresults}
\end{table}

To launch a \sopPFL attack on the iOS platform, a ``malicious'' HTML file must be delivered to the victim app.
We have identified three such channels.
\begin{itemize}

\item[\textbf{Local}] The adversary can design stealthy iOS apps (e.g., the Jekyll app \cite{Jekyll13}) to launch local \sopPFL attacks, because some iOS apps accept external HTML files from other apps through the ``open with'' feature\footnote{Two ``open with'' demos implemented in Dropbox and WeChat are available at \url{http://goo.gl/H7KXeM}.}.
Similarly, Android browsers often use exposed browsing interfaces \cite{FileCross14}.
We notice that local attacks can also be conducted ``remotely'' by leveraging other apps' remote channels.
For example, an adversary first sends an HTML file to the popular WeChat app installed on a victim device.
Since WeChat will not open this type of file, it will ask the user to open the file using another app.
%\TODO, as shown in \myfig \ref{}.
%It is easy for the adversary to motivate the victim to open the file using a vulnerable app.
We find that browser and cloud drive apps are likely to be affected by these local attacks.

\item[\textbf{Web}] An adversary can deliver attack vectors through the web service interfaces of mobile apps.
For example, cloud drive and note apps support file sharing on the web.
An adversary can share an HTML file with a victim via web interfaces.
%When the victim opens the file in the corresponding iOS app, the web-driven \sopPFL attack succeeds.

\item[\textbf{Remote}] Remote \sopPFL attacks are possible for some iOS apps.
The attachment mechanism in email apps is an ideal channel to launch targeted remote \sopPFL attacks.
For example, once a Mail.Ru iOS user opens an attached HTML file from an adversary email, the adversary can steal the victim's private Mail.Ru files remotely.
Similarly, the file sending mechanism in social apps, such as the popular Tencent QQ, can also be exploited.
%These remote \sopPFL attacks are targeted, easy to launch, and very serious.

\end{itemize}

\ifMOST
\else
We have reported these vulnerabilities to their vendors (mostly in early June 2014).
The vendors have acknowledged our reporting and some of them have patched the issues\footnote{Currently developers have to implement application-level defenses (e.g., refusing external files from other apps and disabling JavaScript in the \file domain) to patch their apps, as the underlying iOS engine is still vulnerable.}.
In particular, Mail.Ru and 360 have awarded us with several bug bounties.
For example, Mail.Ru has issued a total of \$1,000 for its two \sopPFL vulnerabilities.
Other vendors, such as Evernote, Baidu, and Tencent, have either listed us in their security halls of fame or given us bug bounty gifts.
\fi

Besides HTML files, we anticipate other types of files for launching \sopPFL attacks.
For example, the commonly used PDF and flash file formats can execute embedded JavaScript codes in a desktop environment~\cite{CVE-2014-0521, FlashOver12}.
\ifMOST
The current mobile systems have limited support for flash and only run basic JavaScript in PDF files.
\else
The current mobile systems have limited support for flash~\cite{iOSFlash, AndroidFlash} and only run basic JavaScript in PDF files \cite{pdfJS}.
\fi
Once these systems are improved, we expect that these new attack vectors will bypass the existing protection.
For example, WeChat is immune to the current \sopPFL attacks by disabling the opening of HTML files. However, it allows opening PDF files to be opened, which could be exploited for future \sopPFL attacks.

%=============
%  aim PFL
%=============
\subsection{AimIFL Vulnerabilities}
\label{sec:aimPFL}

We summarize the \aimPFL vulnerabilities in Table \ref{tab:aimPFL}.
As discussed in Section \ref{sec:aimAtk}, these vulnerabilities can be classified into two types: \aimPFLa and \aimPFLb.

% \TODO talk about how to find them, the method can be extended to test other apps.

\textbf{aimIFL-1 attacks.}
We attempt an \aimPFLa attack via \file on two Android apps, Baidu Browser and On The Road.
We find it difficult to directly load a \file content (e.g., via an HTML iframe) from a web page on Android.
%\footnote{Our testing reveals that this protection is also implemented in modern desktop browsers such as Chrome and Firefox.}.
We thus use an alternative method that asks users to click a \file link embedded in a web page.
\ifMOST
This method is able to exploit On The Road, as the app renders the \file link clickable.
\else
This method is able to exploit On The Road (see \myfig \ref{fig:117go} in Appendix \ref{supporting}), as the app renders the \file link clickable.
\fi
However, similar to desktop browsers, \file links in Baidu Browser are not clickable.
We find that an adversary can entice a victim to long-press the link, allowing Baidu Browser to pop up a dialog that the user can click.
%It is easy to motivate victims to perform this step with an appropriate text indication.
The target file is then rendered and its contents are stolen automatically.
The attack procedure is illustrated in \myfig \ref{fig:baidubrow}.

\begin{table}[t!]
\centering
\vspace{-3ex}
\caption{\small Apps vulnerable to the \aimPFL attacks.}
\ifMOST
\vspace{-2ex}
\fi
\scalebox{
\ifACM
\ifELS
0.9
\else
0.95
\fi
\else
1
\fi
}{
\begin{adjustbox}{center}
\begin{tabular}{ |l | c | }
\hline
Attack Name & Vulnerable Apps  \tabularnewline
\hline
\hline

\aimPFLa via \file & Baidu Browser, On The Road \tabularnewline
\hline
\aimPFLa via \content & 360 Mobile Safe    \tabularnewline
\hline
\multirow{2}{*}{\aimPFLa via \intent} & Yandex and 360 browsers  \tabularnewline
 		      & Baidu Search, Baidu Browser  \tabularnewline
\hline

\multirow{3}{*}{\aimPFLb on Android}  & \texttt{org.easyweb.browser} \tabularnewline
                                      & Internet Browser, Smart Browser \tabularnewline
                                      & Shady Browser, Zirco Browser \tabularnewline
\hline
\aimPFLb on iOS & myVault  \tabularnewline
\hline
\end{tabular}
\end{adjustbox}
}
\ifMOST
\vspace{-1.5ex}
\fi
\label{tab:aimPFL}
\end{table}

\begin{figure}[t!]
\begin{center}
\includegraphics[width=0.48\textwidth]{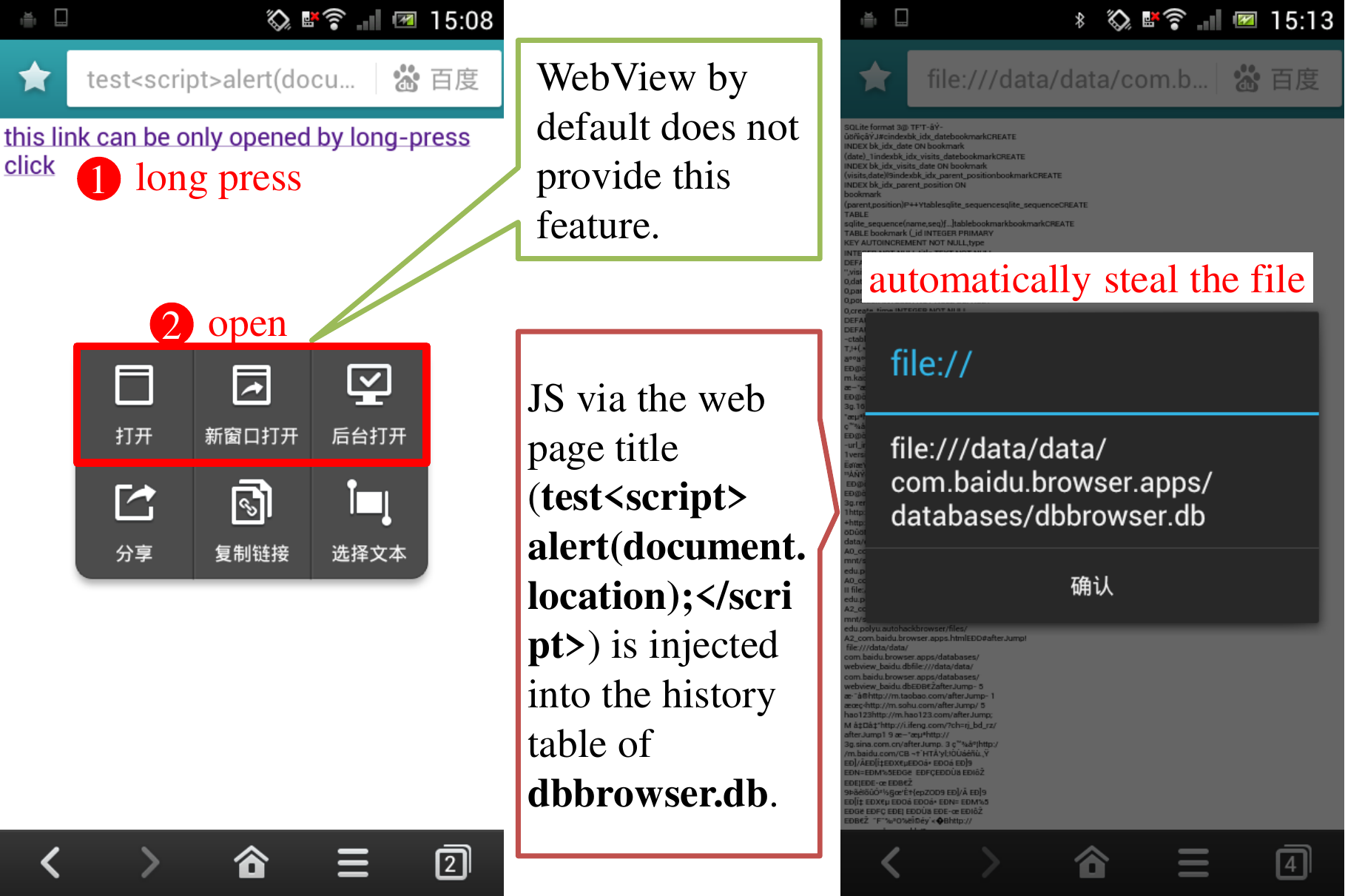}
\end{center}
\vspace{-2ex}
\caption{\small An \aimPFLa attack exploiting Baidu Browser.}
\vspace{-2ex}
\label{fig:baidubrow}
\end{figure}

We find the \content scheme on Android can also be exploited by \aimPFLa attacks.
This scheme is used to retrieve content or data from the corresponding content provider components.
%\cite{ContentUris, ContentProviders}
Surprisingly, we find that a web page can load a local file via the \content scheme if the associated content provider implements the \texttt{openFile(Uri uri, String mode)} API.
Launching \aimPFLa attacks via \content is therefore even easier than \file.
% \TODO Depending on the \texttt{openFile} implementation, not every
\ifMOST
We have successfully launched this attack to remotely exploit 360 Mobile Safe.
\else
We can use this attack to remotely exploit 360 Mobile Safe (see \myfig \ref{fig:360safe} in Appendix \ref{supporting}).
\fi
Because of the seriousness of this exploit, 360 has issued us the highest award in its mobile bug bounty program history.

%\TODO how to find, condition?

\ifPET
\else
\input{360fig}
\fi

Terada \cite{IntentScheme14} points out that the \intent scheme can be used to remotely attack local Android components.
Following this idea, we independently identify several popular browsers and the Baidu Search app that can be exploited by \aimPFLa attacks via \intent.
These victim apps satisfy the following three conditions, which make them exploitable.
\begin{compactitem}
\item They contain \texttt{Intent.parseUri()} to intercept an \intent URI and generate an \texttt{Intent} structure.
  This \texttt{Intent} can invoke any component of the victim, even a private component that has not been exposed to other apps.
  An adversary can thus design a crafted \intent URI to deliver attack vectors to a target component.

\item They include a component that imports external \texttt{Intent} parameters to \texttt{WebView.loadUrl(String url)}.
  An adversary can thus control this component to render an arbitrary URI.

\item The victim component in the last step allows \file access and its JavaScript execution.
  The victim can therefore render a target file via \file and execute its embedded JavaScript codes.
\end{compactitem}

%\TODO steal what kinds of files; JS alert() does not work in some components

%\TODO other browsers, like QQ (no valid), Maxthon (do not spend time)

All the \aimPFLa attacks discussed above affect only Android apps.
We will explain why it is hard to launch \aimPFLa attacks on iOS in Section \ref{sec:implication}.

\textbf{aimIFL-2 attacks.}
Launching the \aimPFLb attacks successfully requires two conditions.
The victim app must be able to load the content of a target file into a WebView-based UI gadget, and the adversary must be able to inject an unauthorized JavaScript into the target file.
%As illustrated in \myfig \ref{fig:aimPFL2}, the bookmark and browsing history files are good candidates of these loaded files.

We use the first condition to facilitate the search for vulnerabilities.
We focus on Android browsers and iOS WebView-based apps in \cite{PhoneGapiOSApp} and inspect their screenshots to choose which apps to install and test.
The \path{WebView.loadDataWithBaseURL} API is useful for locating vulnerable Android browsers, because developers often invoke this API to load the local content and their associated assets.
%\cite{loadDataWith1, loadDataWith2}.
We use this API and its first parameter value (e.g., starting with ``file://'') to search Android browser codes and identify a vulnerable open source Zirco browser.
Interestingly, this vulnerable Zirco design is also used in several other browsers, including ``Browser for Android'' (\texttt{org.easyweb.browser}) that affects around one million users.

One iOS app, myVault, is exploitable by the \aimPFLb attack.
This app allows users to store their private photos, bookmarks, and passwords.
Its bookmark store page is an entry point where a ``malicious'' bookmark can be injected to steal the victim's bookmarks.
Even worse, as iOS does not enforce SOP well on \file, an adversary can therefore steal other sensitive content through a crafted bookmark.

%\TODO \myfig \ref{} shows the attack demos of Zirco Browser and myVault.
%\TODO required or not?

%\TODO Real-world Javascript Injection.
%a malicious QR code.
%mp3, wifi.

%\TODO We anticipate there are other vulnerable apps, as a whole.

%=============
%  cmd PFL
%=============
\subsection{CmdIFL Vulnerabilities}
\label{sec:cmdPFL}

%This section describes the identified \cmdPFL vulnerabilities in popular apps, which demonstrate that the \cmdPFL attack proposed in Section \ref{sec:cmdAtk} is realistic.

Table \ref{tab:cmdPFL} lists the identified \cmdPFL vulnerabilities.
We select command terminal and server apps in Google Play, because these apps are more likely to contain command interpreters than normal apps.
More specifically, we evaluate the top apps, Terminal Emulator and SSHDroid, as they are the most likely to be installed by users with these requirements.
For example, Terminal Emulator for Android has over 10 million installs, whereas the top two terminal apps have less than 0.5 million installs.
%To test the \icmdPFL case, we analyze popular Android apps for the insecure usage of the addJavascriptInterface API. We dynamically evaluate potential remote code execution issues.

\begin{table}[t!]
\centering
\vspace{-3ex}
\caption{\cmdPFL vulnerabilities.}
\ifMOST
\vspace{-2ex}
\fi
\scalebox{
\ifACM
\ifELS
0.78
\else
0.85
\fi
\else
1
\fi
}{
\begin{adjustbox}{center}
%\begin{threeparttable}
\begin{tabular}{ |c | c | c | c| }
\hline
\multirow{2}{*}{Apps} & \multirow{2}{*}{Vulnerability Cause} & Attack & \# of \tabularnewline
 &  & Channel & Installs \tabularnewline
\hline
\hline

Terminal & \multirow{2}{*}{The command component is exposed.}  & \multirow{2}{*}{Local} & \multirow{2}{*}{10M+} \tabularnewline
Emulator &              &                        & \tabularnewline
\cline{1-4}
\multirow{2}{*}{SSHDroid} & The command server is     & Local \&                & \multirow{2}{*}{500K+} \tabularnewline
                          & weakly protected.   & Intranet               & \tabularnewline
\hline

%\multirow{7}{*}{\icmdPFL} & Maxthon       &                         & remote & 10M+* \tabularnewline
%                          & Ali LaiWang   & remote code execution   & remote & 10M+ \tabularnewline
%                          & Ali TianMao   & via the insecure        & MITM   & 40M+ \tabularnewline
%                          & 58.com        & addJavascriptInterface  & MITM   & 20M+ \tabularnewline
%                          & Yahoo Shop JP & API invocation          & MITM   & 100K+ \tabularnewline
%                          & On The Road   &                         & local  & 3M+ \tabularnewline
%\hline

\end{tabular}
%\begin{tablenotes}
%\item [*] The number of installs for the Chinese apps (e.g., Maxthon and Ali LaiWang) are counted in the Baidu App Market because Google Play is blocked in China.
%\end{tablenotes}
%\end{threeparttable}
\end{adjustbox}
}
\label{tab:cmdPFL}
\end{table}

\begin{figure}[t!]
\vspace{-2ex}
\begin{center}
\begin{adjustbox}{center}
\includegraphics[width=0.5\textwidth]{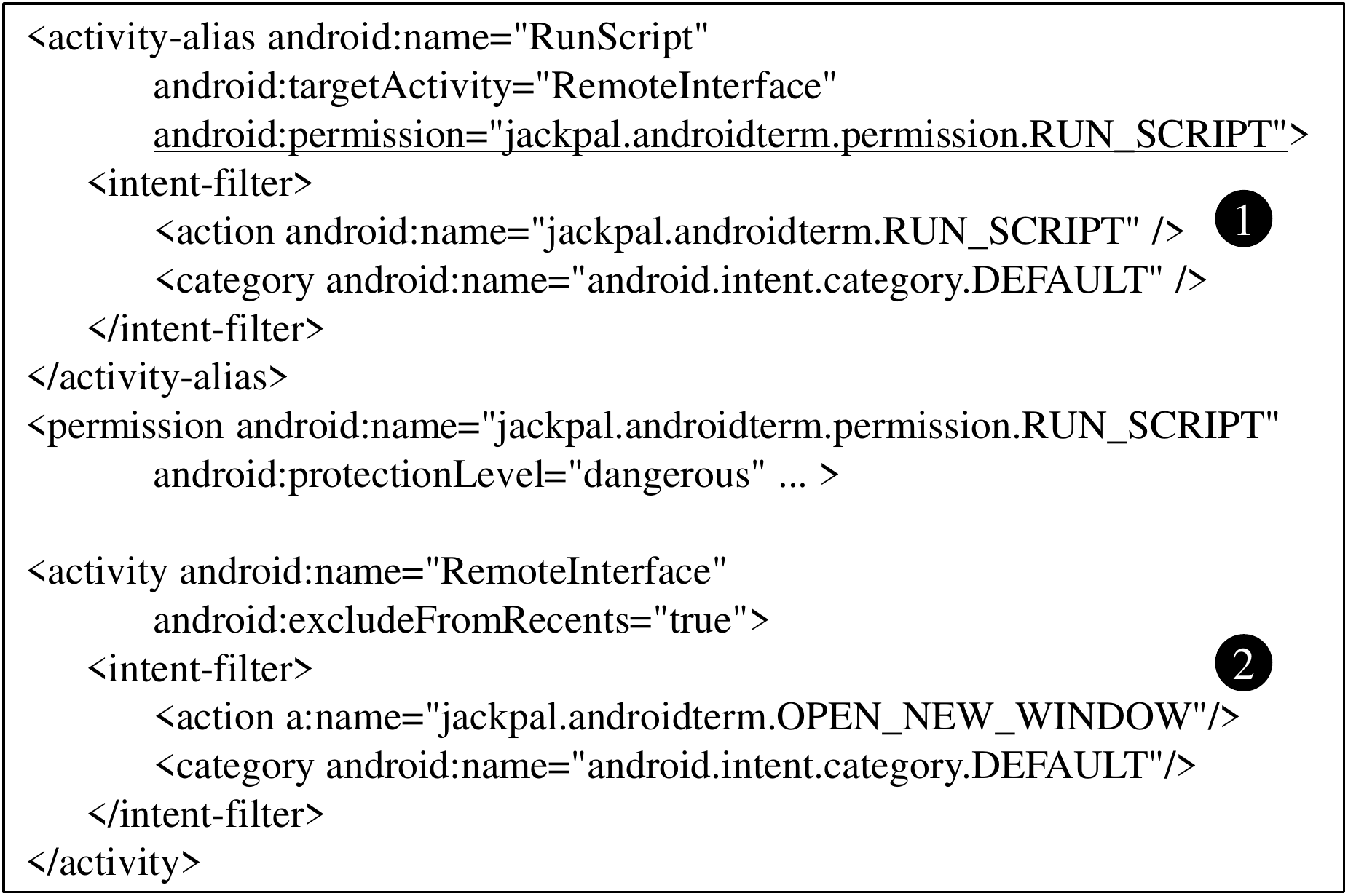}
\end{adjustbox}
\end{center}
\vspace{-3ex}
\caption{Manifest excerpt of Terminal Emulator.}
\vspace{-4ex}
\label{fig:terminal}
\end{figure}

\begin{table*}[ht!]
\centering
\vspace{-3ex}
\caption{\serPFL security weaknesses in the top 10 server apps.}
\ifMOST
\vspace{-2ex}
\fi
\scalebox{
\ifACM
\ifELS
0.82
\else
0.9
\fi
\else
1
\fi
}{
\begin{adjustbox}{center}
\begin{threeparttable}
\begin{tabular}{ |c | c | c | c | c | c | l | c | c | c| }
\hline
         & \multirow{2}{*}{App} & \multirow{2}{*}{App}  &          &      & \multirow{2}{*}{Transmission} &                  & Immune to   & Effective    & \multirow{3}{*}{\# of Installs*}  \tabularnewline
Platform & \multirow{2}{*}{Id}  & \multirow{2}{*}{Name} & Protocol & Port & \multirow{2}{*}{Encryption}   & Authentication   & File Upload & Connection   & \tabularnewline
         &                      &                       &          &      &                               &                  & CSRF        & Alert        & \tabularnewline
\hline
\hline

\multirow{5}{*}{Android} & 1 & AirDroid            & http & 8888 & \x (setting)       & \y (user confirm)  & \y            & \y  & 10M - 50M   \tabularnewline
                         & 2 & WiFi File Transfer  & http & 1234 & \x (setting)       & \x (setting)       & \x            & \x  & 5M - 10M    \tabularnewline
 	                     & 3 & Xender              & http & 6789 & \x                 & \y (four numbers)  & \y            & \x  & 1M - 5M     \tabularnewline
 	                     & 4 & WiFi File Explorer  & http & 8000 & \x                 & \x (setting)       & \RIGHTcircle  & \x  & 1M - 5M \tabularnewline
                  & 5 & \texttt{com.file.transfer} & ftp  & 2121 & \x                 & \x                 & \y            & \x  & 100K - 500K \tabularnewline
\hline

\multirow{5}{*}{iOS}  & 6 & Simple Transfer     & http & 80   & \x                 & \x (setting)       & \y            & \RIGHTcircle & 1,504 Ratings \tabularnewline
 	                  & 7 & Photo Transfer WiFi & http & 8080 & \x                 & \y (six bytes)     & \y            & \x  & 865 Ratings \tabularnewline
 	                  & 8 & WiFi Photo Transfer & http & 15555& \x                 & \x (setting)       & \y            & \x  & 661 Ratings \tabularnewline
 	             & 9 & USB \& Wi-Fi Flash Drive & http & 8080 & \x                 & \x                 & \x            & \x  & 462 Ratings \tabularnewline
                      &10 & Air Transfer        & http & 8080 & \x                 & \x (setting)       & \y            & \x  & 138 Ratings  \tabularnewline
% 	                      & Air Disk Free       & http & 8988 & \x                 & \x                 &    & \x  & 43 Ratings \tabularnewline
% 	                      & Air Drive           & http & 8000 & \x                 & \x                 &    & \x  & 33 Ratings \tabularnewline
\hline

\end{tabular}
\begin{tablenotes}
\item [*] The app install numbers were counted on November 1, 2014. We use rating numbers to estimate the popularity of the iOS apps.
\end{tablenotes}
\end{threeparttable}
\end{adjustbox}
}
\ifMOST
\vspace{-4ex}
\fi
\label{tab:serverPFL}
\end{table*}

Both tested command apps are found to be vulnerable to the \cmdPFL attacks.
A local attack app can execute arbitrary commands using the identity of Terminal Emulator, such as exporting its private files (e.g., command histories and private configuration files) to a public SD card.
The root cause of this vulnerability is that the exposed terminal command component can be invoked by other local apps with arbitrary command parameters.
Interestingly, we find that Terminal Emulator tries to protect its command component with a \texttt{dangerous}-level permission, as shown in part (1) of \myfig~\ref{fig:terminal}.
The rationale behind this design is that all command invocations (via the \texttt{jackpal.androidterm.RUN\_SCRIPT} action) now have the appropriate authorization through the permission mechanism.
Unfortunately, the adversary does not need to touch the protected command proxy component (``RunScript'').
Instead, it directly invokes the underlying command component (``RemoteInterface'' in part (2) of \myfig \ref{fig:terminal}), which is by default exposed by Android's intent filter mechanism \cite{ComDroid11}.
%Consequently, a crafted input (see our demo code in \cite{TermDemo}) can force the ``RemoteInterface'' component to execute commands.
Consequently, a crafted input can force the ``RemoteInterface'' component to execute commands.
We reported this issue and its demo attack code to Terminal Emulator's github page, and helped the open-source community patch the app \cite{Term375, Term374}.

The top command server app, SSHDroid, is vulnerable to the local and Intranet \cmdPFL attacks.
This app works as an SSH server listening to the default port of 22, but it cannot prevent unauthorized connectors.
An adversary does not need to fingerprint SSHDroid, because SSHDroid gives this information to any connector.
SSHDroid only has two user name choices, the ``user'' in the normal case or the ``root'' on a rooted phone.
It also uses the default ``admin'' password if the user does not change it in the settings.
Hence, an adversary has enough pre-knowledge to exploit this app and steal its private files.
More alarmingly, SSHDroid always tries to work as the ``root''.
If it is exploited on a rooted phone, an adversary can execute root commands and steal the private files of all the installed apps.

%\TODO private key, force running by a local app, root

%We demonstrate \icmdPFL vulnerabilities through the Android addJavascriptInterface RCE issue \cite{addJavascriptInterfaceSaga}.
%Industry reports \cite{BingLook, GiltLook, IndeedLook} have reported some affected apps.
%We report more instances to show the pervasiveness of this issue.
%Table \ref{tab:cmdPFL} lists some of the newly identified vulnerable apps.
%They include the popular Maxthon Browser, the Alibaba group's LaiWang and TianMao, etc.
%We can use the remote, man-in-the-middle, and local channels to deliver attack vectors to the vulnerable browsing components, which contain the insecure addJavascriptInterface invocations.
%All of these vulnerable apps can be exploited to leak files.

%\TODO iOS and normal apps. common apps may also contain command interpreters \ldots

%\TODO remote control in iPhone, still little [Implication 3]

%\url{https://play.google.com/store/apps/details?id=com.jrummy.root.browserfree}

%=============
%  server PFL
%=============
\subsection{ServerIFL Vulnerabilities}
\label{sec:serPFL}

We test the top ten server apps from Google Play and the Apple App Store to evaluate the \serPFL vulnerabilities.
Table \ref{tab:serverPFL} summarizes the statistics of their security metrics.
%\TODO introduce how these apps work?
All of the tested apps have at least one security weakness that can be exploited to launch the \serPFL attacks.
%Indeed, most of them (six out of nine tested apps) suffer from two critical security weaknesses (i.e., unencrypted transmission and lack of authentication).
%This causes a very weak adversary can even

Surprisingly, none of these apps provide encrypted transmission between file requesters (e.g., users' desktop browsers) and file servers (i.e., the tested apps).
Eighty percent of the apps do not implement this important security guarantee at all.
This can cause serious consequences in the wireless setting, which is assumed in these apps' user models.
Two apps also provide this functionality which, however, is not enabled by default.
We find that the apps' SSL encryption (when manually enabled in the setting) accepts only self-signed certificates, which causes security warnings in client-side browsers, thus hurting the user experience.

The authentication used in these apps is very weak.
Seven of the ten tested apps do not enforce authentication, including the most popular iOS server app, Simple Transfer.
An Intranet adversary can thus easily send unauthorized file extraction requests.
The apps that conduct authentication still do not have guaranteed security due to the aforementioned lack of encrypted transmission.
An adversary can sniff wireless traffic to obtain the secret information used for authentication or the cookies used for post-authentication transmission.
The secret information used for authentication is generally not strong, such as the four-number verification code used in app \#3 and the six-character password in app \#7.
Brute-force attacks are therefore practical.
For example, to crack Xender's authentication, an adversary only needs to try 10,000 times at most.

%Authentication is also weak, also mention how to break
%\TODO
%how to break authentication?
%
%* sniff plain text
%
%* forget password
%
%* RE cookie generation algorithm
%
%* cookie is unified
%
%* read password from setting files, or just disable it
%
%* CSRF maybe, but mainly for unauthorized deleting files
%
%* airdroid can leak internal IP to external websites via Refer, making remote attacks possible. but require the CSRF
%
%* how can leak airdroid's key via refer?

%remote attack: no anti-CSRF, leak IP address
Remote \serPFL attacks are also possible.
We propose an improved file upload CSRF (cross-site request forgery) attack \cite{fileuploadCSRF} for this purpose.
An adversary uploads an HTML or PDF file with malicious JavaScript codes through the file upload CSRF.
After the victim opens the uploaded file in his desktop browser, the embedded JavaScript runs in the same domain as other target files and can steal arbitrary file content.
We only describe the test results here:
\begin{compactitem}
\item Apps \#2 and \#9 suffer from file upload CSRF, making remote \serPFL attacks possible.

\item Apps \#4, \#5, and \#8 do not support file upload functionalities, making them immune to file upload CSRF. However, the paid version of app \#4 supports file uploading.

\item Apps \#6 and \#7 allow only uploading photo files, making them non-vulnerable.

\item Apps \#3 and \#10 do not support viewing uploaded files. Therefore, the attack vector (i.e., the embedded JavaScript) cannot be executed in the victim's desktop browser.

\item One app, AirDroid, embeds a secret token into each GET/POST request URL. As CSRF cannot obtain this token, the app is safe.
\end{compactitem}

Launching stealthy \serPFL attacks usually requires that victim apps do not have an effective mechanism to detect illegal connections.
Our evaluation reveals that only two apps have such detection capabilities.
AirDroid alerts users to confirm or reject each new incoming connection and breaks the last connection.
This mechanism is effective in preventing stealthy connections, because it is hard for a local attack app to disable or envelop AirDroid alerts.
Simple Transfer displays a ``connected'' UI when it accepts its first connection.
However, it does not further implement multiple-connection detections or warnings.
This weakness allows an adversary to stealthily connect to a victim app after the victim has established its initial connection with a legal user browser.

To launch effective \serPFL attacks, an attacker could fingerprint common server apps in advance.
As shown in Table \ref{tab:serverPFL}, the protocols used in the tested apps are quite indistinguishable (basically HTTP). However, the opened ports have sufficient variability for identifying the apps.
Moreover, for the apps with the same port numbers, an adversary can leverage different HTTP responses to distinguish them.
Once the adversary has constructed a database of fingerprints, it can launch targeted \serPFL attacks on the apps.

%\TODO Vulnerability lab seems find a lot

%\TODO \intent in MeiTuan

%SMS Shell Commander

%app server, file transfer, AirDroid, feedly that own server, ftp server, web server

\section{Android vs iOS}
\label{sec:implication}

Our evaluation reveals four major differences between Android and iOS in terms of the impact of the \name attacks.
We discuss their implications below.

\textbf{Implication 1:}
\textit{The common practice in iOS apps to open (untrusted) files in their own app domain could lead to more pervasive and powerful \sopPFL attacks on iOS than Android.}
%Opening external files within the own app process

\ifMOST
We compare the file-opening behavior in the iOS and Android versions of representative apps in different categories.
We choose two file types, HTML and PDF, for their ability to carry attack vectors.
We find that most of the tested iOS apps open the HTML files by themselves.
\else
According to Table \ref{tab:openCompare}, most of the tested iOS apps open HTML files by themselves.
\fi
In contrast, the corresponding Android versions choose to let dedicated apps (e.g., browsers) handle the HTML files.
The PDF files have similar results.
Opening untrusted files within the app's own domain is thus a common practice in iOS apps, whereas Android apps generally ask dedicated apps to open files.
However, Google Drive and WeChat for iOS also require explicit user actions to open HTML files outside the app. But similar to other iOS apps, they open PDF files internally.
%Another exception, Asana, uses a web link to open files instead of downloading them as local files.

This common practice produces more attack surfaces for iOS apps than their Android versions.
Asking dedicated apps to handle untrusted files is a more secure design, because potential attack vectors are then kept away from the user's private files.
The tested Android apps generally use this practice, which makes them immune to \sopPFL attacks.
Hence, \sopPFL attacks on Android are nearly local attacks that force files opened in \textit{exposed} browsing interfaces, which only affects browser apps and careless apps.
\ifMOST
In contrast, iOS cases are more pervasive and span multiple app categories.
\else
In contrast, iOS cases are more pervasive and span multiple app categories (see Table \ref{tab:iOSresults}).
\fi
They are also more powerful and can be local, web, or remote attacks, as they do not necessarily require locally exposed components.

There are many possible reasons for iOS's this design.
We believe that the lack of flexible data sharing on iOS is an important reason responsible for the apps to open files by themselves.
Indeed, the iOS data sharing involves a non-lightweight process of ``exporting and importing,'' possibly due to the lack of public SD cards and a content URI mechanism, both of which are supported on Android. %\cite{ContentUris}
%We identify the following three causes.
%First, keeping files open within an app may be a UI design choice, as it ensures a consistent user experience.
%Second, iOS lacks a flexible data-sharing mechanism between apps.
%Indeed, iOS data sharing requires the non-lightweight process of ``exporting and importing,'' possibly due to the lack of public SD cards and a content URI mechanism, both of which are supported on Android. %\cite{ContentUris}
%We believe that the lack of flexible data sharing on iOS is an important reason why apps open untrusted files by themselves.
%\ifMOST
%Third, we find that iOS UIWebView has PDF plugin support, whereas Android WebView does not yet contain a PDF plugin.
%\else
%Third, we find that iOS UIWebView has PDF plugin support \cite{iOSPDF}, whereas Android WebView does not yet contain a PDF plugin~\cite{AndroidPDF}.
%\fi
%It is therefore easier for iOS apps to open PDF files than the corresponding Android versions.

\textbf{Implication 2:}
\textit{The randomized app data directory on iOS makes it difficult to conduct the \aimPFLa attacks on iOS.}

The \aimPFLa attacks usually require the knowledge of a full file path.
However, iOS assigns a random directory for each app's data zone, which makes it difficult for a remote attacker to construct the full path of a target file.
Moreover, this iOS randomness is performed at every installation. Therefore, the directory of an app reinstalled on the same phone will be different after each new installation.
An example of a randomized app directory is \path{3570E343-2A5A-484E-BC86-7B3CC611D634}, with the unified path prefix \path{/private/var/mobile/Applications/} (on iOS~7) and \path{/private/var/mobile/Containers/Bundle/Application/} (on iOS 8).

In comparison, Android names app data directories according to the app package name.
An adversary can easily construct the app directory using the pattern \path{/data/data/packagename/}.
As apps generally do not use their own randomness within this directory, it is straightforward to obtain the full file path.
We only find one exception: Firefox uses a random path strategy in its Android app design.
An example of its full file path is \path{/data/data/org.mozilla.firefox/files/mozilla/}\underline{\path{62x7scuo.default}}\texttt{/cookies.sqlite} with the randomized directory underlined.

As randomizing the app directory is useful for thwarting the \aimPFLa attacks on iOS, we recommend the Android developers to use this practice in their own app design.

\textbf{Implication 3:}
\textit{Apple's strict app review prevents iOS apps from executing bash commands. An adversary therefore cannot find targets to launch the \cmdPFL attacks on iOS.}

As stated in \cite{Jekyll13}, Apple has strict regulations for reviewing iOS apps submitted to the App Store.
Apple's app review guidelines \cite{AppleReview} briefly describe many scenarios that can lead to an app rejection.
Among them, \textit{rule 2.8} states:
\ifMOST
\else
\vspace{-1ex}
\fi
\begin{quote}
\textit{Apps that install or launch other executable code will be rejected.}
\end{quote}
\ifMOST
\else
\vspace{-1ex}
\fi

%\url{http://stackoverflow.com/a/1046757/197165}
%\url{http://stackoverflow.com/questions/14170416/executing-a-bash-command-in-an-ios-app}
%\url{http://stackoverflow.com/q/12972802/197165}
%\url{http://www.gamasutra.com/blogs/KerryJones/20130604/193593/iOS\_interpreted\_code.php}
\ifMOST
This rule implies that running interpreted codes (e.g., bash scripts) is forbidden by Apple.
\else
This rule implies that running interpreted codes (e.g., bash scripts) is forbidden by Apple, as discussed by \cite{iPhoneBASH, iOSInterpreted}.
\fi
We thus cannot locate any iOS apps that contain command interpreters, which is a necessary condition for launching the \cmdPFL attacks.
Although a few iOS apps (e.g., ipash ME) claim that they provide command execution for iOS, they actually only mimic the output and do not run the native commands.
They therefore receive customer reviews saying ``It's a fake command line.''

To sum up, this iOS restriction makes it nearly impossible to launch the \cmdPFL attacks on iOS, because there are no suitable app targets in the App Store.

\textbf{Implication 4:}
\textit{iOS generally does not allow background server behavior, which reduces the chance of the \serPFL attacks on iOS.}

%%describe scenario
The success of launching the \serPFL attacks depends on whether the adversary can attack victim apps when the phone screen is off or locked.
If victim apps do not support background servers, then the attack timing window is shortened, thus reducing the chance of a successful \serPFL attack.
The evaluation in Section \ref{sec:serPFL} indicates that iOS server apps usually only work in the foreground.
Of the top five iOS server apps, only Air Transfer can be attacked when the screen is off.
In contrast, all of the top five Android apps support background server behavior and are thus exploitable in the same phone setting.

%%reason
We find that it is not easy for iOS developers to implement background server behavior.
They require advanced tricks~\cite{iOSbackground1, iOSbackground2} and have to worry about violating Apple app review policies.
Thus, developing an app server that can run in the background is uncommon on iOS.

%\textbf{Implication 3:}
%\TODO android render the target files

%\textbf{Implication 4:}
%\textit{private file zone.}

\section{Mitigation Methods}
\label{sec:mitigate}

Application-specific defenses are required to mitigate existing \name risks.
Developers can refer to Section \ref{sec:evaluate} for avoiding the same flaws shown in the existing vulnerable apps.
System flaws in Android and iOS, such as the SOP flaw mentioned in Section \ref{sec:sopAtk}, should be also timely fixed.
Four implications in Section \ref{sec:implication} will be useful to improve both app and system security at different levels.
For example, it is prudent for iOS apps not to open untrusted files in their own app domain and instead to ask dedicated apps to handle them.

We now offer two more suggestions, \textit{NoJS} and \textit{AuthAccess}, to further mitigate \name attacks.
\begin{compactitem}

\item NoJS: disabling JavaScript execution in local schemes to safeguard against the \sopPFL and \aimPFL attacks. The \serPFL attacks based on file upload CSRF can be similarly stopped by opening uploaded files as plain texts.

\item AuthAccess: restricting commands and network requests to access apps' private file zone. Each access should be explicitly authorized by users. By doing so, the \cmdPFL and \serPFL attacks can be mitigated.

\end{compactitem}

Beyond vendors' own ad hoc fixes, a central mitigation solution is desirable.
A possible way is to extend the existing SEAndroid MAC system \cite{SEAndroid13} by leveraging the fine-grained context information to differentiate \name attack requests from normal requests.
Prior works \cite{IPCInspection11, Quire11, Taming12, FlaskDroid13} have shown how to collect and enforce process-related context for tackling local permission leak attacks.
The local \name attacks can be handled in a similar way.
However, it is challenging for context-based enforcement to mitigate the remote \name attacks because remote entities are usually not under defender's control and thus their context cannot be easily obtained.
To address this problem, recently proposed user-driven and content-based access control \cite{UDAC12, ContentBasedIsolation13} may be useful.
We will investigate how to leverage them to develop an enhanced context-based MAC system for the \name attack mitigation in our future work.

\section{Related Work}
\label{sec:related}

\textbf{File leaks in mobile apps.}
Compared with the IFLs studied in this paper, direct file leaks are a more straightforward type of file leak.
\ifMOST
\else
Many direct leaks in mobile apps have been reported \cite{Skype11, Opera11, Lookout11, Applock13, FirefoxLeak13, Outlook14, Evernote14, QQLeak12, YoudaoNote13}.
\fi
Many of these leaks were caused by the setting an insecure (e.g., world-readable) permission for its private files in the apps.
\ifMOST
For example, Opera \cite{Opera11} and Lookout \cite{Lookout11} have made this error.
\else
For example, Skype \cite{Skype11}, Opera \cite{Opera11}, Lookout \cite{Lookout11}, and Firefox~\cite{FirefoxLeak13} have made this mistake.
\fi
On the other hand, the victim app writes sensitive files to public storage (e.g., SD cards and system debug logs).
Outlook \cite{Outlook14} and Evernote \cite{Evernote14} put their users at risk in this way.
The recent SEAndroid MAC system defends against these direct leaks, whereas our \name is still an unsolved threat.
Moreover, these direct cases are just local leaks on Android, whereas we propose multiple forms of remote file leaks across both Android and iOS.
%logcat belongs to the first category

Some \name attack instances have been studied before, but they focus only on their specific problems.
%Our model makes it easier to understand prior \name attacks.
For example, Zhou et al. \cite{ContentScope13} study an attack with exposed Android content providers as confused deputies.
By issuing unauthorized database queries to these components, an adversary can steal victim apps' database files.
This attack is one kind of local \name attacks and only exists on Android.
%To summarize, we can call it the \texttt{cpIFL} or \texttt{dbIFL} attack.
Another example is our previous FileCross attacks \cite{FileCross14}, which belong to the \sopPFL attacks discussed in Section \ref{sec:sopAtk}.
However, it \cite{FileCross14} only shows local \sopPFL attacks on Android, whereas we demonstrate that \sopPFL issues also exist in a number of iOS apps and can be remotely exploited.

It is worth noting that a blog post \cite{DropDrive12} reported two local \sopPFL issues in Dropbox and Google Drive iOS apps over two years ago, but did not mention how to deliver the attack vectors (as we do in Section \ref{sec:sopPFL}).
This blog post used the old iOS systems before the recent iOS 7 and 8 to test the problem and did not show that this issue is widespread in the current iOS ecosystem.
We are also the first to identify its fundamental cause: the legacy web SOP does not adequately cover the local schemes.
%download File attack, as opposed to the \aimPFL attack. but the deputy is the download component.
%most of them focus on Android, and local
Compared with all of these isolated \name studies, we are the first to systematically study both local and remote \name attacks across Android and iOS.

%NDSS13_SEAndroid
%Oakland12_UDAC(BPP)
%\TODO 3. mitigation solutions are for direct file leaks

%OSDI14_CleanOS
%\TODO encryptions\ldots

%=======  ECV  =======
%
%ISC10_PrivilegeEscalationAttacks
%MobiSys11_ComDroid
%UsenixSec11_IPCInspection
%UsenixSec11_Quire
%NDSS12_Woodpecker
%NDSS12_Towards_Taming_Privilege-Escalation_Attacks_on_Android
%WiSec12_DroidChecker
%DSN12_IntercomponentCommunication *
%SPSM12_IntraComDroid
%CCS12_CHEX
%NDSS13_ContentScope
%NDSS2013_SEAndroid
%ESORICS12_Sorbet *
%TR13_PermissionFlow *
%AsiaCCS13_DroidAlarm_short
%UsenixSec13_ICCEpicc
%UsenixSec13_FlaskDroid
%CCS13_UnauthorizedCrossOrigin
%ISSA13_APSET *
%CCS13_CustomRom
%
%=======  ECV  =======
\textbf{Confused deputy problems on mobile.}
The \name attack is a class of the general confused deputy problem \cite{Confused88}.
A number of previous works have studied the permission-related confused deputy problem on Android, called permission leak or privilege escalation \cite{ISC10_Privilege}.
They have proposed detection systems based on control- and data-flow analysis, including ComDroid \cite{ComDroid11}, Woodpecker \cite{Woodpecker12}, CHEX \cite{CHEX12}, ECVDetector~\cite{ECVDetector14}, Epicc \cite{ICCEpicc13}, and SEFA~\cite{CustomRom13}.
Some Android app analysis frameworks, such as FlowDroid \cite{FlowDroid14} and Amandroid~\cite{Amandroid14}, can be extended to detect this problem.
However, it is difficult for these static tools to analyze the \name vulnerabilities, because most of the \name attacks do not have explicit vulnerable code patterns.
Using dynamic analysis tools (e.g., \cite{TaintDroid10, DroidScope12, SMVHunter14, Brahmastra14}) to construct automatic \name detectors is therefore desirable.
We leave this for our future work.
Furthermore, none of the aforementioned studies identify confused deputy problems on iOS.
But as shown in our \name attacks, it is equally, if not more, important to develop detection tools for iOS.
%\TODO Most previous mobile vulnerabilities studied are local attacks. We broaden this

\ifPET
\else
To mitigate the privilege escalation, researchers have devoted efforts to design several solutions \cite{IPCInspection11, Quire11, Taming12}.
Access control frameworks, such as FlaskDroid \cite{FlaskDroid13} and ASM \cite{ASM14}, could be also plugged with corresponding defense policies.
They address the problem from different aspects, but all require extracting context information---inter-component communication caller-callee relationship---to defeat untrusted requests attacking confused deputies.
Such context-based approaches is mainly effective to thwart local confused deputy attacks.
It is hard for them to stop remote attacks (e.g., our \aimPFL and \serPFL attacks), because remote entities are usually not under defender's control (i.e., no suitable context could be extracted from them).
To bypass this fundamental limitation, the recent user-driven and content-based access control \cite{UDAC12, ContentBasedIsolation13} might be new directions.
We will investigate how to leverage them to mitigate our \name attacks in the further research.
\fi

\ifPET
\textbf{Mobile browser security.}
\else
\textbf{Mobile browser security and SOP.}
\fi
%ACSAC11_AttackWebView
%CCS13_UnauthorizedCrossOrigin
%WISA13_BifocalsWebView(BP)
%NDSS14_NoFrak
%WiSec14_NativeWrap
%ISC14_FileCross
%CCS14_CodeInjection
Our \sopPFL and \aimPFL attacks are related to mobile browser security.
Related works have studied the threats to Android WebView \cite{WebViewAtk11, Bifocals13, NativeWrap14, FileCross14}, the security risks in HTML5-based mobile libraries \cite{NoFrak14} and apps \cite{CodeInjection14}, and unauthorized origin crossing in several popular Android and iOS apps \cite{CrossOrigin13}.
Differently, our focus is file leaks via vulnerable browser components.
Only the aforementioned work \cite{FileCross14} shares the same goal as our study.
%Our findings in Section \ref{sec:sopPFL} and \ref{sec:aimPFL} significantly complement the previous work~\cite{FileCross14}.
In addition, a prior work \cite{CodeInjection14} injects unauthorized JavaScripts into HTML5-based apps, and their technique is similar to our \aimPFL attacks.
However, as their goal is to compromise the victim website's online credentials (instead of our goal of stealing local files), they do not need to overcome our challenge of launching local file-stolen attacks from a web origin due to the SOP restriction.
Moreover, our \aimPFL attacks apply to all mobile apps that contain browser components, rather than just the HTML5-based apps shown in \cite{CodeInjection14}.
%ACSAC11\_WebView: still the exposed JS objects, because no returned file object like in content provider
%JS event interception?
%
%but hard to steal files remotely

\ifPET
\else
In Section \ref{sec:sopAtk}, we call for the standardization of enhanced SOP (that considers the ``path'' ingredient into the current tuple of scheme, domain, and port) for local schemes.
We find some modern browsers have adopted this principle, e.g., the full support in Chrome and partial support in Firefox \cite{fileSOPfox}.
Unfortunately, no formal standard has been drafted or even no previous efforts have been made to call for such standardization, to the best of our knowledge.
The legacy SOP was invented years ago to protect web resources at that time, so it is hard to thwart the attacks under new threat models.
For example, two enhanced SOPs were proposed to thwart dynamic pharming attacks \cite{LockedSOP07} and DNS rebinding attacks \cite{eSOP13}, respectively.
We believe the enhance SOP for local schemes is also necessary.
\fi

%TODO CVE flash file SOP
%beside iOS, other windows, firefox mobile systems. as well as the desktop SOP

%\noindent \textbf{Smartphone privacy leakage.}
%%a lot of papers
%The file leak is one serious class of the generic privacy leakage.
%Many works
%
%
%hard to detect
%
%
%traditional privacy leak methods are well known, but file leaks especially IFLs are lesser studied in the literature.

\section{Conclusion}
\label{sec:conclude}

In this paper, we systematically studied indirect file leaks (IFLs) in mobile applications.
In particular, we devised four new \name attacks that exploit browser interfaces, command interpreters, and embedded app servers to leak private files from popular apps.
Unlike the previous attacks that work only on Android, we demonstrated that these IFLs (three of them) can affect both Android and iOS.
The vulnerable apps include Evernote, QQ, and Mail.Ru on iOS, and Baidu Browser, 360 Mobile Safe, and Terminal Emulator on Android.
Moreover, we showed that our four \name attacks can be launched remotely, without implanting malicious apps in victim's smartphones.
This remote attack capability significantly increases the impact of the \name attacks.
Finally, we analyzed the differences between Android and iOS in terms of the \name attacks' impacts and proposed several methods to mitigate the attacks.

\textbf{Acknowledgements}.
We thank all three anonymous reviewers for their helpful comments.
This work was partially supported by a grant (ref. no. ITS/073/12) from the Innovation Technology Fund in Hong Kong.

\textbf{Additional materials}.
We will provide supplementary materials, such as detailed vulnerability reports, at this link (\url{https://daoyuan14.github.io/pp/most15.html}).

\bibliographystyle{abbrv}
\balance
{\small\bibliography{ieeeMain}}
%{\scriptsize\bibliography{file}}

\ifMOST
\else
\newpage
\ifPET
\appendix
\input{appendix}
\fi
\fi

\balance

\end{document}